\documentclass[slac_one]{revtex4}
\usepackage{graphicx}
\usepackage{fancyhdr}
\usepackage{feynmp}
\newcommand{\mrm}[1]{\mathrm{#1}}

\newcommand{\pT}{{p_\perp}}

\newcommand{\MeV}{~\mathrm{MeV}}
\newcommand{\GeV}{\ensuremath{~\mathrm{GeV}}}
\newcommand{\g}{\mathrm{g}}
\newcommand{\tq}{\mathrm{t}}

\newcommand{\tbar}{\bar{\mathrm{t}}}

\newcommand{\go}   {\tilde{\g}}

\begin{document}

\title{{\small{2005 ALCPG \& ILC Workshops - Snowmass,
U.S.A.}}\\ 
\vspace{12pt}
QCD Radiation in the Production of High-$\hat{s}$ Final States
} 

%

\author{P.~Skands (speaker)}
\affiliation{Theoretical Physics Dept., Fermi National Accelerator Laboratory,
             Batavia, USA}
\author{T.~Plehn}
\affiliation{Heisenberg Fellow, Max Planck Institute for Physics, Munich, Germany}
\author{D.~Rainwater}
\affiliation{Marshak Fellow, Dept.of Physics and Astronomy, University
  of Rochester, Rochester, USA}

\begin{abstract}
In the production of very heavy final states --- high Mandelstam
$\hat{s}$ --- extra QCD radiation can play a significant role. By
comparing several different parton shower approximations to results 
obtained with fixed--order perturbation theory, we quantify the 
degree to which these approaches agree (or disagree), focussing on 
initial state radiation above $\pT=50\GeV$, for top pair production at
the Tevatron and at the LHC, and for SUSY pair production at the LHC. 
Special attention is paid to ambiguities associated with 
the choice of the maximum value of the ordering variable 
in parton shower models. 
\end{abstract}
\maketitle
\section{Introduction}
Hadron colliders like the Tevatron and the LHC are ideal machines for
producing strongly interacting states. 
To the extent that any new such
state is related to a WIMP type explanation for the dark
matter problem, a decay chain generally exists by which the coloured 
particle decays to jets plus missing $E_\perp$ (a stable WIMP), 
possibly accompanied by one or more leptons. 
A prime example of such a type of collider phenomenology
is given by supersymmetry (SUSY), which simultaneously provides not
only a possible solution to the dark matter problem, 
but also the maximal space--time symmetry, 
an elegant solution to the ultraviolet instability of the Higgs mass
(the hierarchy problem), radiative breaking of electroweak symmetry, 
and gauge coupling unification. 
As such, the phenomenology of supersymmetric particle
production at colliders has been a topic of ever increasing interest
over the last few decades. 

In the case of supersymmetry, 
the relevant hadron collider processes are pair production of squarks and
gluinos \cite{Dawson:1983fw,Beenakker:1996ch,Beenakker:1997ut},
followed by cascade decays down to the 
Lightest Supersymmetric Particle (LSP), usually the
$\tilde{\chi}^0_1$, which, if $R$--parity is conserved, has WIMP-like
properties and escapes the detector unobserved.

Since gluino decays produce more jets than
squark decays, the jet multiplicity in the event can be used to
separate squark- and gluino-enriched samples. Also, 
since the LSP escapes detection, the decay
kinematics cannot be fully reconstructed. Instead, the masses involved
in the decay chain can be determined from kinematic edges in the
invariant mass spectra of the observed particles and jets
\cite{Bachacou:1999zb,Kawagoe:2004rz}. As such, any extra jet activity
in the event, e.g.\ from QCD bremsstrahlung, will introduce a source
of combinatorial error, hence our interest in studying the production
of extra QCD radiation in association with these processes. Note that
most of our conclusions should be applicable, in general, to the production of
any high-mass strongly interacting states. 

At currently accessible collider energies, 
the QCD coupling strength, $\alpha_s$, is large,
ranging from the non--pertubative, at scales $\le \Lambda_{\mrm{QCD}}\sim
100\MeV$, to $\alpha_s\sim\mathcal{O}(0.1)$ at the highest energy 
scales relevant for the LHC. 
Thus, even in the perturbative domain of QCD, higher--order
corrections are typically large, 
and one has to exercise caution in estimating the uncertainties
coming from uncalculated contributions, whether these be due to higher orders 
or to other sources.  

In this paper, we focus especially on the production of very heavy
states, or more precisely processes with large factorisation scales,
at the Tevatron and at the LHC. Following some general comments in
Section \ref{sec:meps}, we take a closer look at three specific high-mass 
processes in Section \ref{sec:results}: $\tq\tbar$ at the Tevatron,
$\tq\tbar$ at the LHC, and SUSY pair production at the LHC. Finally,
in Section \ref{sec:conclusions} we present a summary and
outlook. This report closely follows the studies reported in
\cite{Plehn:2005cq}. 

\section{Fixed Order vs.\ Resummation \label{sec:meps}}
There are essentially two widespread and somewhat complementary
methods to calculate perturbative QCD amplitudes; 
fixed--order matrix elements and resummation / parton shower approaches. 

Computing fixed--order matrix elements, we 
include the full dynamical, helicity, interference, and phase space
structure, up to a given order in the coupling constants (here only
$\alpha_s$ will be relevant). This is a procedure which
by now is highly automated (for a brief overview, see
e.g.\ \cite{Skands:2005hd}). 
We can also include virtual corrections, i.e.\ quantum loops;
although complete results beyond one loop (NLO) are still scarce. Two
problems occur in this approach, firstly that the complexity rapidly 
increases, both as a function of the number of legs and, even more
rapidly, as a function of the number of loops, and secondly;
bremsstrahlung corrections contain singularities
which, for soft and collinear radiation, renders a 
truncation of the perturbative series unstable at any fixed order in
certain regions of phase space.

On the other hand, in the collinear limit we are in fact able
to sum the perturbative series to infinite order in the coupling
constant, hence curing the truncation problem. 
Parton showers are examples of such approaches. The main
virtue is that these descriptions should work well at low $\pT$, 
and (squared) amplitudes 
for final states with an arbitrary number of partons can be
built out of relatively simple expressions. In addition, due 
to a combination of factorisation and 
universality, these descriptions can also be more easily macthed onto
hadronisation models than fixed-order approaches. Nonetheless, since
we are working in a particular \emph{limit} of QCD, uncertainties and
ambiguities appear as soon as we try to extrapolate away from that
limit. Common problems for parton shower models include 
a simplified treatment of helicity structure, ambiguities in the
size of the radiation phase space, and unknown corrections from
contributions which vanish in the collinear limit.


Especially for hard radiation, large differences may exist between
different shower algorithms. In \textsc{Pythia}
\cite{Sjostrand:2000wi,Sjostrand:2003wg}, 
two qualitatively different shower algorithms are implemented:
one $Q^2$-ordered~\cite{Sjostrand:1985xi,Bengtsson:1986gz,Bengtsson:1986hr,Bengtsson:1986et}, and the other 
$\pT$-ordered~\cite{Sjostrand:2004ef}. 
Due to the large final state masses and since we force the tops and
gluinos to be stable here, we are mainly exploring the
properties of the initial-state showers, for which the crucial parameter
is the starting scale of the shower. Nominally, this scale is identical to the
factorization scale, $\mu_F$, where the parton densities are convoluted with
the matrix elements.

For the $\pT$-ordered shower, $\mu_F$ can be used directly as the
maximum $\pT$.  Below, we refer to this choice as the $\pT$-ordered `wimpy
shower'. Allowing the parton 
shower to populate the full phase space, with the maximum
$\pT_j=\sqrt{s}/2$, regardless of $\mu_F$, we refer to as
the $\pT$-ordered `power shower' ---  strictly speaking in
conflict with the factorization assumption, but with 
interesting phenomenological consequences, as we shall see.

The case of a $Q^2$-ordered shower is not so simple.  The starting
scale here is $Q^2_{\mrm{max}} = \min\left(C \mu_F^2, s\right)$, where
$C\ge 1$ parameterizes the translation from $\pT^2$ to $Q^2$.  We refer
to $C = 1$ as the $Q^2$-ordered wimpy shower, $C=4$ as
Tune~A~\cite{tunea,Field:2005sa}, and $C\to\infty$ as power shower --- with the
same caveat concerning factorization as for the $\pT$-ordered version.

For the fixed-order results, we use tree-level matrix elements as
implemented in the new supersymmetric
version~\cite{smg} of the event generator
MadEvent~\cite{Stelzer:1994ta,Maltoni:2002qb}.  The factorization
scale is set to the average final 
state mass, as is the renormalization scale for the heavy pair.  The 
renormalization scale for additional jet radiation is $\pT_j$. 

\section{Results \label{sec:results}}
In Tab.~\ref{tab:njet} we first show the inclusive production cross sections
obtained in fixed-order perturbation theory  
for top and gluino production plus zero to two hard jets with
$\pT_j>50\GeV$ and $\Delta R_{jj}>0.4$ (for the
gluinos we use the SUSY parameter point SPS1a~\cite{Allanach:2002nj} 
with $m_{\go}=608\GeV$). 
We also compare with a toy top quark $T$ with a mass
of 600\GeV, to test the universality of the behaviour. 

\begin{table}[t]
\begin{center}
\begin{tabular}{|r|c|rrrr|}
 \hline
  & & \multicolumn{1}{c}{Tevatron} & \multicolumn{3}{c|}{LHC}\\
  & $\sigma_{\rm tot} [\rm{pb}]$
  & $\tq\tbar$~~
  & $\tq\tbar$ 
  & $\go \go$ 
  & $T\bar{T}$ \\
  \hline \hline
  $\pT_j>50\GeV$  &$\sigma_{0j}$& 5.13~~ & 461 & 4.83 & 1.30\\
                 &$\sigma_{1j}$& 0.45~~ & 273 & 5.90 & 1.50\\
                 &$\sigma_{2j}$& 0.04~~ & 127 & 4.17 & 1.21\\
  \hline
\end{tabular}
\caption{Cross sections for the production of $\tq\tbar$ pairs at the
  Tevatron and at the LHC, and for gluinos (in SPS1a) 
as well as a toy-model $600\GeV$
top quark $T$ at the LHC. We show fixed-order matrix
element results with 0,1,2 additional hard jets, with $p^{\rm
  min}_{\perp,j}>50\GeV$, $|y_j|<5$ and $R_{jj}>0.4$.} 
\label{tab:njet}
\end{center}
\end{table}

For top pairs at the Tevatron, we see that extra hard jets are not
alarmingly frequent; 
for 50\GeV\ jets, the $\tq\tbar j$ cross section is roughly
10\% of the inclusive $\tq\tbar$ one. Going to $\tq\tbar$ at the LHC, 
the production of extra jets increases dramatically, 
not only due to the increase in phase space, but also to
the more gluon-dominated and hence more active initial state, and to
the top quarks here being produced at larger relative
velocities. A more detailed study of radiation in top events 
can be found in \cite{Orr:1994na,Orr:1996pe}. 

For the production of 50\GeV\ jets in association with even
higher-mass objects, here gluinos and our heavy toy quarks $T$, we see
that fixed-order perturbation theory reaches its limit; 
the $1j$ and $2j$ cross sections are not noticeably smaller than the $0j$
inclusive one, hence a truncation at any finite order is not likely to
give a trustworthy answer.Only by considering jets much harder
than 50\GeV\ would the stability of the pertubative series be recovered.

What is happening is that the soft and collinear singularities of QCD
radiation are introducing logarithmic enhancements of the form
$(\alpha_s \log^2(Q_{\mrm{hard}}^2/\pT_j^2))^N$ at all orders $N$. If
the argument of the logarithm becomes large enough to
counterbalance the $\alpha_s$ suppression, then these corrections
cannot be neglected, and a truncated calculation will give a
meaningless answer. 

\begin{figure}[t]
\begin{center}\vspace*{-7mm}
\hspace*{-7mm}\includegraphics*[scale=0.5]{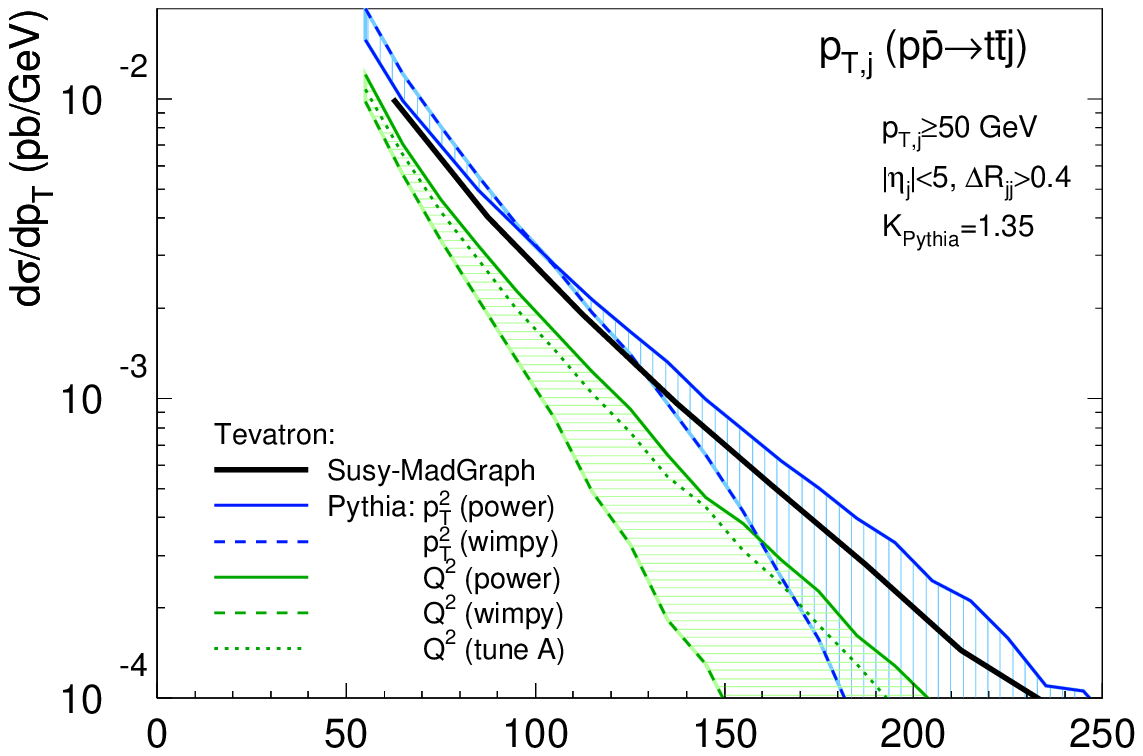}
\hspace*{-7mm}\hspace*{-7mm}\includegraphics*[scale=0.5]{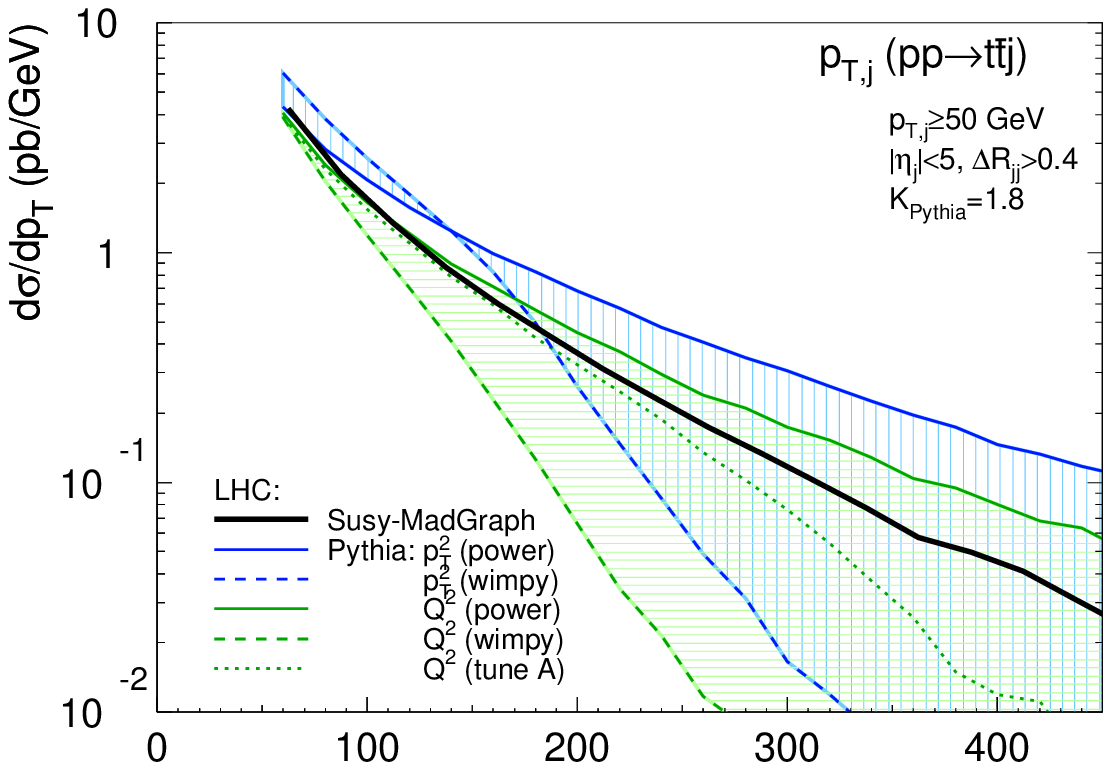}
\hspace*{-7mm}\hspace*{-7mm}\includegraphics*[scale=0.5]{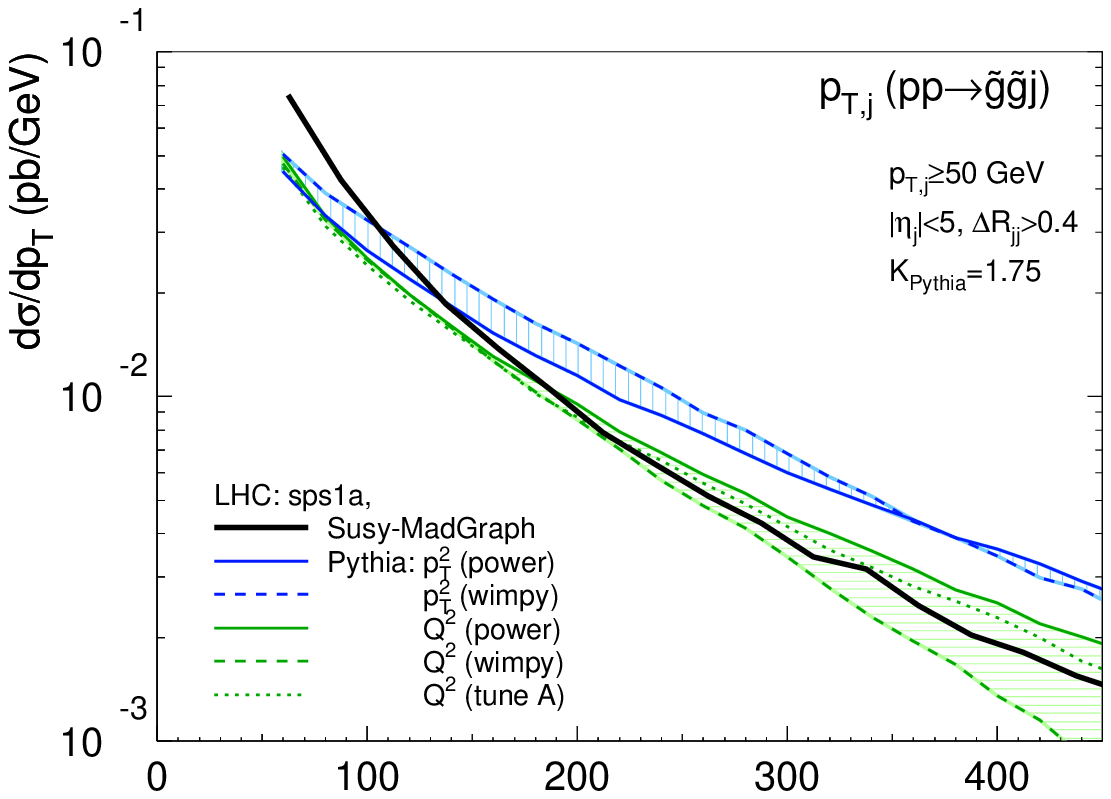}
\hspace*{-7mm}\vspace*{-7mm}
\caption{Jet $\pT$ distributions for $\tq\tbar+j$ at the 
  Tevatron and at the LHC, and for $\go\go+j$ (in SPS1a) at the
  LHC. Comparison between fixed--order matrix elements (solid thick line) 
and parton shower models (blue: $\pT$-ordered, green: $Q^2$-ordered)
  for wimpy (thin dashed) and power (thin solid) showers. The dotted
  green line indicates `Tune A' of
  the $Q^2$-ordered shower.}
\label{fig:jetrates}
\end{center}
\end{figure}

In Fig.~\ref{fig:jetrates} we show the jet transverse-momentum
distributions for $\tq\tbar+j$ at the Tevatron and the LHC, and for 
$\go\go+j$ at the LHC. We compare the fixed-order calculation (thick 
black line) to the 5 different parton shower models described above. 

For $\tq\tbar$ at the Tevatron, we concluded above that there is no
reason not to trust the fixed-order results (thick solid line), 
for the range of $\pT_j$
values we consider here. The interesting feature, then, is that the
power showers (thin blue and green lines) 
do surprisingly well. While the agreement with
the fixed-order result is not perfect, one could have
expected a collinear-based approximation to do worse. The
wimpy showers (dashed blue and green), 
on the other hand, drop off rapidly around the
factorisation scale, due to the presence of the explicit phase space
cutoff, as has also previously been noted e.g.\ for Drell--Yan production
\cite{Mrenna:2003if,Giele:1990vh,Seymour:1994df,Huston:2004yp,Miu:1998ju}. 
(The crossover at $\sim100\GeV$ between the two
$\pT$-ordered showers illustrates the effect of changing the
renormalization scale from $\pT/2$ in the wimpy shower to $3\pT$ in
the power shower.  The $Q^2$-ordered showers all use $\mu_R=\pT_j$.)

For $\tq\tbar$ at the LHC, we have already noted
that the total jet rates are larger, hence the slope of the $\pT$
spectrum is here much gentler than at the Tevatron (note the change in
$\pT$ scale). This means that a correct description of harder jets
becomes relatively more important, and hence the cataclysmic drop of
the wimpy showers potentially more serious. Still, the 
asymptotic slopes of the
power showers are not greatly different from that of the matrix
element. If anything, the rate of hard jets is overestimated by the
power showers (probably indicating that the neglected terms, which can
be roughly classified as interference terms, are negative). 

The tendency of the power showers to overestimate the hard jet tail 
is  seen also in the last plot, $\go\go+j$ production. 
In addition, the drop of the
wimpy showers is here not as catastrophic, ironically due to the
larger masses involved. With a factorisation scale of $\sim 600\GeV$,
the presence or not of a phase space cutoff at that scale does not
produce a large impact for the $\pT$ range we consider here. On the
other hand, at low $\pT$, we see that the fixed-order approximation
starts breaking down already around 100\GeV, while the parton shower
results converge to a common value.

\section{Conclusions \label{sec:conclusions}}
We show that fixed-order QCD calculations predict a large number of hard
jets associated with the production of heavy coloured states at the
LHC. This additional radiation should be 
taken into account in studies of the separation of
squark and gluino event samples, as well as  for cascade decay reconstruction.

Comparisons of jet $\pT$ spectra 
between matrix elements and parton showers show
that, for the radiation of one extra jet, 
conventional (wimpy) parton showers with a phase space
cutoff at the factorization scale give a reasonable approximation up
to $\pT_j \sim \mu_F/2$, above which they rapidly break down. 
Removing the phase space cut tends to yield somewhat harder radiation spectra
than produced by the matrix elements.The tendency of these `power
showers' 
to overestimate
the rate of hard jets may be useful for future studies for which parton
shower approximations are applied in hard regions of phase space. 
This is especially relevant for processes where 
the correct higher higher order matrix elements are
not known, such as would be the case for many exotic BSM physics
scenarios.

For fairly soft jets, we see that in the production of high-mass
gluinos the breakdown of fixed-order perturbation theory caused by
logarithmic corrections can occur already at jet transverse momenta of
as high as 100~GeV.

\section*{Acknowledgements}
We express our gratitude to the organisers of the Snowmass
2005 workshop, for creating a stimulating and rewarding atmosphere,
and for the kind invitation to present our work there. We would also 
like to thank T.~Sj\"ostrand and P.~Richardson for enlightening
discussions and comments on the manuscript.  This research
  was supported in part by the U.S. Department of Energy under grant
  Nos. DE-FG02-91ER40685 and DE-AC02-76CH03000.

\bibliography{qcdrad2}

\begin{thebibliography}{10}

\bibitem{Dawson:1983fw}
S.~Dawson, E.~Eichten, and C.~Quigg.
\newblock {\em Phys. Rev.}, D31:1581, 1985.

\bibitem{Beenakker:1996ch}
W.~Beenakker et~al.
\newblock {\em Nucl. Phys.}, B492:51--103, 1997.

\bibitem{Beenakker:1997ut}
W.~Beenakker et~al.
\newblock {\em Nucl. Phys.}, B515:3--14, 1998.

\bibitem{Bachacou:1999zb}
H.~Bachacou, I.~Hinchliffe, and F.~E. Paige.
\newblock {\em Phys. Rev.}, D62:015009, 2000.

\bibitem{Kawagoe:2004rz}
K.~Kawagoe, M.~M. Nojiri, and G.~Polesello.
\newblock {\em Phys. Rev.}, D71:035008, 2005.

\bibitem{Plehn:2005cq}
T.~Plehn, D.~Rainwater, and P.~Skands.
\newblock 2005.
\newblock hep-ph/0510144.

\bibitem{Skands:2005hd}
P.~Z. Skands.
\newblock {QCD} (and) event generators.
\newblock In {\em Proceedings of 13th International Workshop on Deep Inelastic
  Scattering (DIS 05), Madison, Wisconsin, 27 Apr - 1 May 2005}, 2005.
\newblock hep-ph/0507129.

\bibitem{Sjostrand:2000wi}
T.~Sj{\"o}strand et~al.
\newblock {\em Comput. Phys. Commun.}, 135:238--259, 2001.

\bibitem{Sjostrand:2003wg}
T.~Sj{\"o}strand et~al.
\newblock 2003.
\newblock hep-ph/0308153.

\bibitem{Sjostrand:1985xi}
T.~Sj{\"o}strand.
\newblock {\em Phys. Lett.}, B157:321, 1985.

\bibitem{Bengtsson:1986gz}
M.~Bengtsson, T.~Sj{\"o}strand, and M.~van Zijl.
\newblock {\em Z. Phys.}, C32:67, 1986.

\bibitem{Bengtsson:1986hr}
M.~Bengtsson and T.~Sj{\"o}strand.
\newblock {\em Phys. Lett.}, B185:435, 1987.

\bibitem{Bengtsson:1986et}
M.~Bengtsson and T.~Sj{\"o}strand.
\newblock {\em Nucl. Phys.}, B289:810, 1987.

\bibitem{Sjostrand:2004ef}
T.~Sj{\"o}strand and P.~Z. Skands.
\newblock {\em Eur. Phys. J.}, C39:129--154, 2005.

\bibitem{tunea}
R.~D. Field.
\newblock hep-ph/0201192 CDF Note 6403; further recent talks available from
  webpage {\texttt{http://www.phys.ufl.edu/}$\sim$\texttt{rfield/cdf/}}.

\bibitem{Field:2005sa}
R.~Field and R.~C. Group.
\newblock 2005.
\newblock hep-ph/0510198.

\bibitem{smg}
K.~Hagiwara et~al.
\newblock in preparation.

\bibitem{Stelzer:1994ta}
T.~Stelzer and W.~F. Long.
\newblock {\em Comput. Phys. Commun.}, 81:357--371, 1994.

\bibitem{Maltoni:2002qb}
F.~Maltoni and T.~Stelzer.
\newblock {\em JHEP}, 02:027, 2003.

\bibitem{Allanach:2002nj}
B.~C. Allanach et~al.
\newblock {\em Eur. Phys. J.}, C25:113--123, 2002.

\bibitem{Orr:1994na}
L.~H. Orr, T.~Stelzer, and W.~J. Stirling.
\newblock {\em Phys. Rev.}, D52:124--132, 1995.

\bibitem{Orr:1996pe}
L.~H. Orr, T.~Stelzer, and W.~J. Stirling.
\newblock {\em Phys. Rev.}, D56:446--450, 1997.

\bibitem{Mrenna:2003if}
S.~Mrenna and P.~Richardson.
\newblock {\em JHEP}, 05:040, 2004.

\bibitem{Giele:1990vh}
W.~T. Giele et~al.
\newblock W boson plus multijets at hadron colliders: {HERWIG} parton showers
  versus exact matrix elements.
\newblock Contribution to Proc. of 1990 Summer Study on High Energy Physics:
  Research Directions for the Decade, Snowmass, CO, Jun 25 - Jul 13, 1990.

\bibitem{Seymour:1994df}
M.~H. Seymour.
\newblock {\em Comp. Phys. Commun.}, 90:95--101, 1995.

\bibitem{Huston:2004yp}
J.~Huston et~al.
\newblock Resummation and shower studies.
\newblock 2004.
\newblock in "The QCD/SM Working Group: Summary Report" [hep-ph/0403100], 3rd
  Les Houches Workshop: Physics at TeV Colliders, Les Houches, France, 26 May -
  6 Jun 2003. hep-ph/0401145.

\bibitem{Miu:1998ju}
G.~Miu and T.~Sj{\"o}strand.
\newblock {\em Phys. Lett.}, B449:313--320, 1999.

\end{thebibliography}

\end{document}